# Time Reversal of Arbitrary Photonic Temporal Modes via Nonlinear Optical Frequency Conversion

Michael G Raymer[1*], Dileep V Reddy[1], Steven J van Enk[1] and Colin J McKinstrie[2]

*1 Department of Physics and Center for Optical, Molecular and Quantum Science, University of Oregon, Eugene, Oregon 97403, USA*

*2 Huawei Technologies, 400 Crossing Boulevard, Bridgewater, New Jersey 08807, USA*

\* raymer@uoregon.edu

**Abstract**
Single-photon wave packets can carry quantum information between nodes of a quantum network. An important general operation in photon-based quantum information systems is "blind" reversal of a photon's temporal wave-packet envelope, that is, the ability to reverse an envelope without knowing the temporal state of the photon. We present an all-optical means for doing so, using nonlinear-optical frequency conversion driven by a short pump pulse. The process used may be sum-frequency generation or four-wave Bragg scattering. This scheme allows for quantum operations such as a temporal-mode parity sorter. We also verify that the scheme works for arbitrary states (not only single-photon ones) of an unknown wave packet.

**1. Introduction**
    Single-photon wave packets can carry quantum information between nodes of a quantum network for distributed quantum-information processing. For efficient transfer, the shapes of the wave packets need to be finely controlled. For example, the temporal shape emitted by a single quantum emitter (atom, quantum dot, etc.) in free space is not the shape that leads to efficient absorption by a receiving quantum absorber, even if the emitter and absorber are physically identical. For efficient absorption, the envelope of the emitted wave packet must be time reversed. [1,2,[3]] Optimal input and output modes of optical cavities have a similar time-reversed relationship, [4,5] as has been demonstrated experimentally for weak classical pulses [6] and for single-photon wave packets. [7]
    Although quantum mechanics is formally time-reversal symmetric, there is no unitary transformation or corresponding physical device operation that can implement full time reversal of a quantum state. [8,9,2] In quantum optics, such an operation would temporally reverse the complex envelope of an optical wave packet and at the same time conjugate its phase, that is $\mathcal{E}(\mathbf{r},t) \to \mathcal{E}^*(\mathbf{r},-t)$. It is known that optical phase conjugation by four-wave mixing can cause such a phase transformation with or without reversal of the envelope [10,11,12], but only at the cost of adding quantum noise (spontaneous photons) to the mode of interest, making the transformation non-unitary. [13,14,15,16] The added noise makes phase conjugation ineffective for implementing quantum-information protocols.

This paper proposes an all-optical operation that reverses only a wave packet's envelope function and not its phase. The method functions without knowing the wave packet; thus we call it "blind." Such an operation can in principle be unitary and thus noise free, as needed in quantum information science.

It is known that reflection by an optical cavity designed to have its resonant frequency and cavity decay time matching those of the quantum emitter reverses the envelope function of an incident wave packet generated by that emitter [6]. But this does not represent a universal of "blind" operation, as it requires that the particular shape of the optical wave packet be known in advance, and fails if the wave packet does not match the cavity. This prevents the technique being used to implement quantum logic operations in a state space of temporal shapes (modes)—an important capability in quantum information science [17]

In addition to noiseless waveform manipulation, in quantum networks there is also a need for noiseless frequency conversion (FC), which together with pulse shaping can enable complete temporal and spectral "impedance" matching between quantum emitters and absorbers that are not identical (for example, a rubidium vapor and a GaAs quantum dot), having different carrier frequencies and decay times. [18,19,20,21] Such frequency conversion has been demonstrated for single photons using nonlinear three-wave mixing to implement sum-frequency generation [22] while maintaining single-photon number statistics [23] and even quantum entanglement [24] with other physical systems. For FC between frequencies separated by smaller differences than can be accommodated by sum- or difference-frequency generation, four-wave mixing has been demonstrated for converting the frequency of single photons. [25] It is known that FC using three- or four-wave mixing can be engineered to be highly waveform selective. [26,27] Ideally, only a single, given TM is frequency converted, while all temporally orthogonal TMs are not. The resulting operation, called a "quantum pulse gate," enables a new framework for quantum information science, based on manipulation of single-photon waveforms and carrier frequencies in a higher-dimensional (qudit) state space. [17]

In this paper we show theoretically that frequency conversion by three- or four-wave mixing can be engineered to implement noiseless envelope reversal in a manner that is waveform *nonselective*, which enables quantum information operations such as time reversing a superposition of envelope shapes without knowledge of the state. Within certain bandwidth constraints (to be specified), it operates on the envelope as $f(t) \rightarrow f(-t)$, that is, it reverses the envelope but does not conjugate the phase. In this regard the time-reversal frequency conversion process is similar to a time lens, but with the difference that it also converts the carrier frequency from one band to another. The new scheme provides a single unifying platform that can implement both frequency conversion and noiseless time reversal of arbitrary (that is, unknown) envelopes.

An alternative method that can reverse an unknown, arbitrary-shaped envelope (within a given spectral band) without adding quantum noise is the so-called time lens. [28,29,30] The time lens uses fast modulation of the refractive index to temporally vary the phase of a light pulse, analogous to the spatial variation of phase imposed by a lens. As a single lens spatially inverts an image in the image plane, the temporal "image" of the incident pulse is time reversed. The same operation can also stretch (magnify) or compress (demagnify) the waveform, leaving the carrier frequency unchanged. [31] Relatively small shifts of carrier frequencies (up to 100s of GHz) can be implemented using the same class of phase modulators used in the time lens. [32] The time lens reshapes a wave packet, but is limited in its ability for frequency conversion compared to the nonlinear-optical method. The presently proposed scheme can implement envelope reversal and large-reach frequency conversion in a single device, creating a useful tool for quantum information networks.

Optical processes that use inhomogeneously broadened atomic media to implement time reversal are photon echoes and quantum memories based on controlled reversible inhomogeneous broadening (CRIB), also known as gradient echo memory (GEM). [33,34,35,36] They are in principle noiseless [37], but extraneous quantum noise associated with atomic or molecular fluctuations and spontaneous emission, as well as the bandwidth and wavelength constraints imposed by the use of specific resonant materials, may render these approaches less flexible for use in quantum information systems than the all-optical schemes such as the time lens and the presently proposed method.

## 2. Time Reversal Considerations

First, consider the general quantum theoretical problem of time reversing an optical pulse. In a one-dimensional model, we restrict our interest to a frequency band denoted $\Omega$ centered at $\omega_0$ with bandwidth $B$. (We don't need this restriction here, but it will prove useful later.) The forward-traveling "photon field" operator (scaled so that $\hat{A}^{(-)}\hat{A}^{(+)}$ is a photon flux) in this band has positive-frequency part:

$$\hat{A}^{(+)}(z,t) = \frac{1}{2\pi}\int_\Omega d\omega\, e^{-i\omega t}\, e^{i\beta(\omega)z}\, \hat{a}(\omega) \qquad (1)$$

where $\beta(\omega) = |\omega|/c$ is the propagation constant (wavenumber) in free space and the single-frequency creation operators satisfy the commutator $[\hat{a}(\omega), \hat{a}^\dagger(\omega')] = 2\pi\delta(\omega - \omega')$.

A pure state of a given temporal mode (wave packet), denoted by the index $j$ within the spectral band $\Omega$, can be expressed in terms of the photon creation operator corresponding to that temporal mode [38,39]:

$$\hat{a}_j^\dagger = \frac{1}{2\pi}\int_\Omega d\omega\, \Psi_j(\omega)\hat{a}^\dagger(\omega) \qquad (2)$$

If the set of spectral amplitudes $\{\Psi_j(\omega)\}$ form an orthonormal set under inner product $(2\pi)^{-1}\int d\omega\, \Psi_j^*(\omega)\Psi_k(\omega)$, then the resulting set of temporal mode (TM) operators obey the discrete boson commutator $[\hat{a}_j, \hat{a}_k^\dagger] = \delta_{jk}$, indicating that different TMs correspond to distinct, discrete degrees of freedom. A single-photon state of a given TM is

$$\left|1_j\right\rangle = \hat{a}_j^\dagger\left|vac\right\rangle = \frac{1}{2\pi}\int_\Omega d\omega\, \Psi_j(\omega)\hat{a}^\dagger(\omega)\left|vac\right\rangle, \qquad (3)$$

and an arbitrary state of the TM can be expressed as $F(\hat{a}_j^\dagger)\left|vac\right\rangle$, where $F$ is some normalizable function. The corresponding temporal modes are given by the spatial-temporal amplitudes:

$$\begin{aligned}\psi_j(z,t) &= \left\langle vac\right|\hat{A}^{(+)}(z,t)\left|1_j\right\rangle \\ &= \frac{1}{2\pi}\int_\Omega d\omega\, e^{-i\omega t}\, e^{i\beta(\omega)z}\, \Psi_j(\omega)\end{aligned} \qquad (4)$$

which form a discrete, orthogonal set under an inner product involving integration over time $t$ or longitudinal distance $z$.

Full time reversal (better named *motion reversal*) corresponds to replacing the temporal amplitude by [8, 9] (where * means complex conjugation)

$$\psi_{j,TR}(z,t) = \psi_j^*(z,-t) , \tag{5}$$

which is equivalent to making the replacements:

$$\Psi_j(\omega) \to \Psi_j^*(\omega) \tag{6}$$

and $\exp[i\beta(\omega)z] \to \exp[-i\beta(\omega)z]$. Full time reversal is seen to be *antilinear*, defined for an operator $\hat{\Pi}$ in quantum theory as $\hat{\Pi}(c\Psi) = c^* \hat{\Pi}(\Psi)$. [8,9] Therefore, no physical process can implement it. To illustrate this point, the time reversal operator applied to one half of an entangled state can produce a state described by a density matrix that is unphysical, evidenced by having at least one negative eigenvalue.

Phase conjugation by four-wave mixing can perform an operation that looks like full time reversal of a signal mode [10, 12], but with added spontaneous noise arising from coupling to an extra degree of freedom—the idler mode. [15]

In contrast to full time reversal, time reversing only the envelope of an optical pulse can in general be carried out by a physical process described by a unitary transformation without any added noise. Consider the temporal mode as a product of an optical carrier at $\omega_0$ and a slowly varying envelope $\varphi_j(z,t)$:

$$\psi_j(z,t) = e^{-i\omega_0 t} \varphi_j(z,t) \tag{7}$$

The temporal envelope function is related to the spectral amplitude by (defining $v = \omega - \omega_0$):

$$\varphi_j(z,t) = \frac{1}{2\pi} \int_{0_A} dv\, e^{-ivt} e^{i\beta(\omega_0+v)z} \Psi_j(\omega_0+v) \tag{8}$$

where $0_A$ indicates a "base band" centered at zero. Envelope reversal (and reversing the direction of propagation) corresponds to $\varphi_j(z,t) \to \varphi_j(z,-t)$, or in the frequency domain

$$\Psi_j(\omega_0+v) \to \Psi_j(\omega_0-v) \tag{9}$$

and $\exp[i\beta(\omega_0+v)z] \to \exp[-i\beta(\omega_0-v)z]$. This transformation is unitary and linear (does not involve complex conjugation of the state).

### 3. Envelope Reversal – Perturbative Theory

Consider an initial temporal mode in the spectral band centered at frequency $\omega_s$. The state of this TM (single-photon, squeezed, coherent, etc.) is arbitrary and unknown. The goal is to time reverse the TM while converting it to a band with a new center frequency $\omega_r$, otherwise preserving the state. The process proposed below, driven by a pump field centered at frequency $\omega_p$, reverses the TM's envelope but not its direction of propagation, which can be accomplished trivially by reflection from an ordinary mirror.

We assume the two signal bands are connected via a nonlinear optical three-wave-mixing process driven by the strong, nondepleting pump field, such that the band centers are related by $\omega_r = \omega_s + \omega_p$. The process is known as sum-frequency generation (SFG). We assume that competing processes, such as dissipative loss, cross-phase modulation, and higher-order dispersion, are all negligible. Denote group slowness (inverse group velocity $1/v_{gj}$) of light in band $\Omega_\eta$ by $\beta_\eta' = \partial_\omega \beta(\omega)$, evaluated at band center $\omega = \omega_\eta$, where now $\beta(\omega)$ accounts for the linear dispersion of the medium. (Four-wave mixing could also be used, as discussed later, but we focus on three-wave mixing for clarity.)

We find that to achieve envelope reversal it is sufficient to have the group slownesses ordered as $\beta_s' > \beta_p' > \beta_r'$ (or the inverse), that is, one of the signals travels faster than the pump, while the other signal travels slower than the pump, and to have the medium long enough so that pulses with unequal slownesses pass completely through one another. We show in the Appendix A that this condition can be satisfied for realistic media. Then the medium acts like an infinitely long one, leading to exact conservation of momentum among the fields, as explained below.

We denote the (positive-frequency) photon annihilation operators for propagating fields in the three bands by

$$\hat{A}_\eta^{(+)}(z,t) = \frac{1}{2\pi} \int_{\Omega_\eta} d\omega \, e^{-i\omega t} \, e^{i\beta(\omega)z} \, \hat{a}_\eta(z,\omega) \;, \quad (\eta = s,r,p) \tag{10}$$

where $\Omega_s$, $\Omega_r$ and $\Omega_p$ denote the relevant spectral bands, and $[\hat{a}_\eta(z,\omega), \hat{a}_\mu^\dagger(z,\omega')] = 2\pi\delta(\omega-\omega')\delta_{\eta\mu}$. The dominant term in the interaction Hamiltonian is:

$$H = \int_{-L/2}^{L/2} dz \, \chi^{(2)} \hat{A}_p^{(+)}(z,t) \hat{A}_s^{(+)}(z,t) \hat{A}_r^{(-)}(z,t) + ha \;, \tag{11}$$

where $\chi^{(2)}$ is proportional to the second-order nonlinear susceptibility and *ha* is the Hermitian adjoint. To first order, the evolution operator is $\hat{U} \approx 1 + \hat{U}^{(1)}$, where

$$\begin{aligned}\hat{U}^{(1)} &= -\frac{i}{\hbar} \int_{-\infty}^{\infty} dt \, H \\ &= -\frac{i}{\hbar} \chi^{(2)} \int_{\Omega_r} \frac{d\omega'}{2\pi} \int_{\Omega_s} \frac{d\omega}{2\pi} \, \Phi(\omega',\omega) \, \hat{a}_p(\omega'-\omega)\hat{a}_s(\omega)\hat{a}_r^\dagger(\omega') - ha\end{aligned} \tag{12}$$

and the phase-matching function is

$$\Phi(\omega',\omega) = \int_{-L/2}^{L/2} dz\, e^{i\Delta\beta(\omega',\omega)z} , \qquad (13)$$

where $L$ is the medium's length. The phase mismatch is approximated, neglecting higher-order dispersion, as

$$\Delta\beta(\omega',\omega) = \beta(\omega'-\omega) - \beta(\omega') + \beta(\omega) \\ \approx \sigma_s \nu - \sigma_r \nu' \qquad (14)$$

where $\sigma_s = \beta'_s - \beta'_p$, $\sigma_r = \beta'_r - \beta'_p$, $\omega = \omega_s + \nu$, $\omega' = \omega_r + \nu'$, and we assumed the zeroth-order terms sum to zero due to either birefringent phase matching or periodic poling to achieve quasi-phase matching. Then the phase matching function evaluates to $\pi L D(\nu' - \sigma_s \nu / \sigma_r)$, where $D(x)$ is a unity-normalized function with maximum value along the line $\nu' = \sigma_s \nu / \sigma_r$ and width in the perpendicular direction in the $(\nu',\nu)$ space equal roughly to $\sigma_r c / L$. The evolution operator $\hat{U}^{(1)}$ is then (with $C = \chi^{(2)} \pi L / \hbar$)

$$-iC \int_{0_r} \frac{d\nu'}{2\pi} \int_{0_s} \frac{d\nu}{2\pi} D(\nu' - m\nu)\, \hat{a}_p(\omega_p + \nu' - \nu)\hat{a}_s(\omega_s + \nu)\hat{a}_r^\dagger(\omega_r + \nu') - ha \qquad (15)$$

where $m = \sigma_s / \sigma_r$ and where $0_\eta$ indicates a base band centered at zero. Say the spectral bandwidths of the pump and the two signals are $B_p, B_s, B_r$. Then the conditions needed for $D(\nu' - m\nu)$ to act like a delta function are $2\pi / (\beta'_p L) \ll B_p$, $2\pi / (|\sigma_s| L) \ll B_s$, $2\pi / (|\sigma_r| L) \ll B_r$. This is equivalent to requiring the medium to be long enough so the three pulses can begin being spatially nonoverlapping and pass completely through one another while in the medium. In this limit, the evolution operator is equal to

$$-iC \int_{0_s} \frac{d\nu}{2\pi} \hat{a}_p(\omega_p + [m-1]\nu)\hat{a}_s(\omega_s + \nu)\hat{a}_r^\dagger(\omega_r + m\nu) - ha , \qquad (16)$$

In the case $\sigma_s = \sigma_r$ ($m = 1$), Eq.(16) shows that the field $\hat{a}_s(\omega_s + \nu)$ exchanges photons with the field $\hat{a}_r^\dagger(\omega_r + \nu)$, and there is no envelope reversal. In contrast, for $\sigma_s = -\sigma_r$ ($m = -1$), Eq.(16) shows that $\hat{a}_s(\omega_s + \nu)$ exchanges photons with $\hat{a}_r^\dagger(\omega_r - \nu)$, in agreement with temporal-mode envelope reversal as described by Eq.(9). This supports our claim that to achieve envelope reversal it is necessary to have one of the signals travel faster than the pump, while the other signal travels slower than the pump, and to have the medium long enough so that the pulses pass completely through one another.

In the more general case that $m$ is negative but not necessarily equal to -1, temporal reversal occurs with either a stretching or compressing effect, by a factor $|m|$, which we call the *spectral magnification*. Spectral compression for frequency conversion has been observed in experiments. [40]

To prove the result explicitly for a single-photon state, and to see what requirements are placed on the state of the pump field, we consider an initial state:

$$|\Psi\rangle_{in} = \int_{0_s} \frac{d\nu}{2\pi} \tilde{\varphi}(\nu) \hat{a}_s^\dagger(\omega_s + \nu) |vac\rangle_{s,r} \otimes |\alpha\rangle_p \tag{17}$$

The *s* signal is in a single-photon TM state with spectral amplitude $\tilde{\varphi}(\nu)$ that peaks near $\nu = 0$, the *r* signal is in vacuum, and the pump is in a coherent state $|\alpha\rangle_p$. In the case $\sigma_s = -\sigma_r$, the output state (within the perturbative approximation) is

$$|\Psi\rangle_{out} = \hat{U} |\Psi\rangle_{in} \approx \sqrt{1-\varepsilon^2} |\Psi\rangle_{in} + \varepsilon |\Psi\rangle^{(1)} \tag{18}$$

where $\varepsilon^2 \ll 1$ is the probability the photon has been frequency converted. It is shown in Appendix B that

$$|\Psi\rangle^{(1)} = N \int_{0_s} \frac{d\nu}{2\pi} \tilde{\phi}_p(\omega_p - [m-1]\nu) \tilde{\varphi}(-\nu) \hat{a}_r^\dagger(\omega_r - m\nu) |vac\rangle_{s,r} \otimes |\alpha\rangle_p \tag{19}$$

where $\tilde{\phi}_p(\omega)$ is the (non-operator) spectral amplitude of the pump's coherent state, square-normalized to unity, and $N$ is an overall normalization coefficient.

Consider the case $\sigma_s = -\sigma_r$ ($m = -1$). Then

$$|\Psi\rangle^{(1)} = N \int_{0_s} \frac{d\nu}{2\pi} \tilde{\phi}_p(\omega_p + 2\nu) \tilde{\varphi}(-\nu) \hat{a}_r^\dagger(\omega_r + \nu) |vac\rangle_{s,r} \otimes |\alpha\rangle_p \tag{20}$$

Now we see explicitly that the spectral amplitude of the generated *r* signal, $\tilde{\varphi}(-\nu)$, is the spectrally mirrored version of the spectral amplitude of the *s* signal, $\tilde{\varphi}(\nu)$. In order to have perfect envelope reversal of the signal, the pump spectrum $\tilde{\varphi}_p(\omega)$ must be constant across the signal's spectrum (shifted to the pump's band for comparison), satisfied if the pump is sufficiently broad band, $B_p \gg B_s, B_r$. That is, the pump pulse must be much shorter in duration than the signal pulses.

Equation (19) also shows that in first-order perturbation theory a pump field in a coherent state does not get entangled with the signal fields (because the coherent state is an eigenstate of $\hat{a}_p$). In first-order perturbation theory, a pump field in any other type of pure state will become entangled (because it's *not* an eigenstate of $\hat{a}_p$). In higher-order perturbation theory even a pump field in a coherent state gets entangled with the signals (because it's not an eigenstate of $\hat{a}_p^\dagger$). However, for a large-amplitude pump field that entanglement is small (because the coherent state is "almost" an eigenstate of $\hat{a}_p^\dagger$. More precisely, $\hat{a}_p^\dagger |\alpha\rangle_p$ has a large overlap with $|\alpha\rangle_p$).

This section has shown from general considerations the conditions required on the pump field and dispersion properties to achieve envelope reversal by FC. We also have seen that, in first-order perturbation theory, the signal fields do not become entangled with the pump field if the pump starts in a coherent state. We also argued it is justified to treat a coherent-state pump as a classical field even in the nonperturbative regime, as in usually done in the FC literature. We shall do so in the following sections.

**4. Envelope Reversal – Nonperturbative Theory**

The above leaves open two questions: Does envelope reversal by frequency conversion work for high conversion efficiencies, and does it work for arbitrary (but low-photon-number) states of the input signal? The answers to both are yes, as we now show.

Above we showed that sufficient conditions for envelope reversal are: (i) the bandwidth relation $B_p \gg B_s, B_r$, (ii) an effectively infinite length of medium, so that $2\pi/(\beta'_p L) \ll B_p$, $2\pi/(|\sigma_s| L) \ll B_s$, $2\pi/(|\sigma_r| L) \ll B_r$, where $\sigma_s = \beta'_s - \beta'_p$ and $\sigma_r = \beta'_r - \beta'_p$, (iii) phase matching for the sum-frequency generation process $\omega_s + \omega_p \to \omega_r$, and (iv) higher-order dispersion can be neglected.

The pump is assumed to be strong, with propagating coherent-state amplitude $A_p(t - \beta'_p z)$, assumed to be unchanging as a function of $t - \beta'_p z$. The Heisenberg equations of motion for the signal field operators are [41]

$$\begin{aligned}(\partial_z + \beta'_r \partial_t) \hat{A}_r^{(+)}(z,t) &= i\gamma A_p(t - \beta'_p z) \hat{A}_s^{(+)}(z,t) \\ (\partial_z + \beta'_s \partial_t) \hat{A}_s^{(+)}(z,t) &= i\gamma A_p^*(t - \beta'_p z) \hat{A}_r^{(+)}(z,t)\end{aligned} \quad (21)$$

It is convenient to define a new effective "local time" variable defined by the pump's velocity, as $t_{LT} \doteq t - \beta'_p z$. (Local time is like a collection of clocks strewn along the path, which are progressively delayed according to their location, like time zones on earth.) In terms of this variable, the pump $A_p(t_{LT})$ is stationary, that is, we are representing the propagation in the "frame" moving along with the pump, and in this frame one signal *appears* to move to the right (increasing *z*) and the other to the left (decreasing *z*). We assume the three pulses are timed so they intersect maximally somewhere within the medium.

Now, for notational convenience, we drop the subscript on $t_{LT}$, and write the equations of motion using local time:

$$\begin{aligned}(\partial_z + \sigma_r \partial_t) \hat{A}_r^{(+)}(z,t) &= i\gamma A_p(t) \hat{A}_s^{(+)}(z,t) \\ (\partial_z + \sigma_s \partial_t) \hat{A}_s^{(+)}(z,t) &= i\gamma A_p^*(t) \hat{A}_r^{(+)}(z,t)\end{aligned} \quad (22)$$

The material-dependent coupling constant $\gamma$ (assumed real and positive) is proportional to the nonlinear susceptibility $\chi^{(2)}$.

We next show that, for a short pump pulse, these equations predict that both the amplitude and phase time dependence of the signal's envelope are reversed upon frequency conversion, while the phase of the carrier wave is not conjugated, consistent with noise-free operation.

## 5. Analytical Solution

Because the equations of motion (22) are linear in the signal operators, they may be solved as if they were non-operator equations. In fact, the input-output relations derived thereby can be viewed as "classical" mode transformations, in this case for temporal modes. This will allow predicting output states for arbitrary (but weak) input states.

Under the stated conditions, an excellent approximation to the exact solution can be found, which is verified numerically below. Choose a local time interval $[t_0, t_1]$ such that at the boundaries of this interaction window all three pulses are non-overlapping. As in Fig.1, assume the signal $\hat{A}_s^{(+)}(z,t)$ is slower than the pump and enters this window from the "left," and $\hat{A}_r^{(+)}(z,t)$ is faster than the pump and enters this window from the "right" (i.e., $m < 0$). In the figure, local time is on the horizontal axis and distance increases in the downward vertical direction; therefore the envelope of a slower pulse will move progressively "rightward" as $z$ increases, while the envelope of a faster pulse will appear to move progressively "leftward" as $z$ increases. (Think of the amplitudes as those that would be recorded by a detector placed at a fixed $z$ value.)

Using a "ray tracing" description, a field value $\hat{A}_s^{(+)}(0,t)$ in the $\omega_s$ pulse enters the diagram at local time $t$, indicated by a black dot, then propagates freely, as indicated by the (red) dashed line, enters the interaction window at space-time point $(z_0, t_0)$, where $z_0 = (t_0 - t)/|\sigma_s|$, and exits the diagram at local time $t + |\sigma_s| L$. And a field value $\hat{A}_r^{(+)}(0,t_2)$ in the $\omega_r$ pulse enters the diagram at local time $t_2$, propagates freely, indicated by a (blue) dotted line, enters the interaction window at space-time point $(z_0, t_1)$, where $z_0 = (t_2 - t_1)/|\sigma_r|$, and exits the diagram at $t_2 - |\sigma_r| L$. In the pump region the rays interact; some portion of the signal becomes the idler and vice versa. Equating these two expressions for $z_0$ yields the relation between $t$ and $t_2$, that is, $t_2 = t_1 + M t_0 - M t$, where $M = |\sigma_r / \sigma_s| = 1/|m|$ is the *temporal magnification*, expressing stretching or compressing of the output pulses relative to their input counterparts. In the limit of a short pump pulse, we can approximate $t_0 \approx t_1 \approx t_p$, where $t_p$ is the time at the center of the pump pulse. Choosing $t_p = 0$ gives $t_2 = -M t$, clearly showing time reversal and magnification.

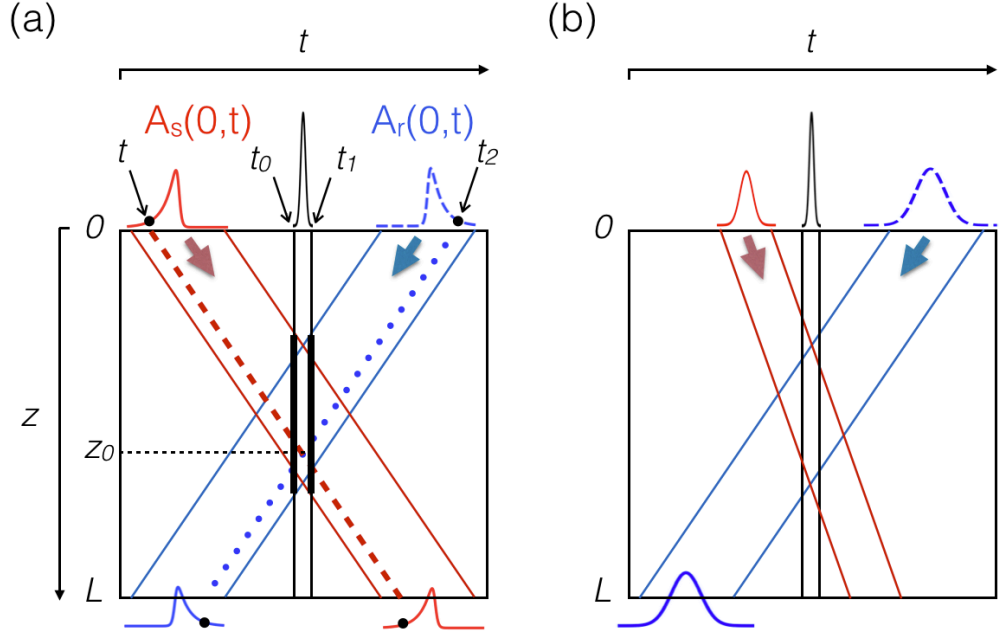

Figure 1 – Ray pictures illustrating envelope time reversal process for typical input pulse shapes, with the $A_r$ field initially in the vacuum state (dashed curve). Distance is $z$ and local time is $t$. The (red) input signal $A_s$ "collides" with the short pump pulse (shown in the pump's moving frame), creating a frequency-converted (blue) signal $A_r$. In (a) the temporal magnification equals 1 and in (b) it is greater than 1. Note that $A_s$ emerges at $z = L$ at the latest time, indicating it travels slowest in the medium, consistent with $\sigma_s > 0$, $\sigma_r < 0$.

With a very short pump pulse, the local interaction between spatial segments of the three pulses occurs in a sufficiently short time interval that spatial propagation has little effect within each segment. Therefore, we can neglect the spatial $z$ derivatives in Eq.(22). Then, defining an effective time variable in terms of the pump function (assumed real):

$$\varepsilon(t) \equiv \int_{t_0}^{t} A_p(t') dt' , \qquad (23)$$

we arrive at:

$$\partial_\varepsilon \hat{A}_r^{(+)}(t) = i(\gamma/\sigma_r)\hat{A}_s^{(+)}(t)$$
$$\partial_\varepsilon \hat{A}_s^{(+)}(t) = i(\gamma/\sigma_s)\hat{A}_r^{(+)}(t) \qquad (24)$$

where both $\sigma_r$ and $\sigma_s$ can be positive or negative.

The initial (input) values depend implicitly on the $z$ value, $z_0$, defining the "rays" along which the solutions propagate. Denoting the values of the two input fields at the boundaries of the interaction window along a particular pair of crossing (dashed) rays by $\hat{A}_s^{(+)}(t_0)$ and $\hat{A}_r^{(+)}(t_1)$, the solutions at the outputs of the interaction window, for the case $\sigma_s > 0$, $\sigma_r = -|\sigma_r|$, are:

$$\hat{A}_r^{(+)}(t_0) = \tau A_r^{(+)}(t_1) + i\rho\sqrt{|\sigma_s/\sigma_r|}\; A_s^{(+)}(t_0)$$
$$A_s^{(+)}(t_1) = \tau A_s^{(+)}(t_0) + i\rho\sqrt{|\sigma_r/\sigma_s|}\; A_r^{(+)}(t_1) \quad (25)$$

where $\tau = \text{sech}(\gamma\varepsilon_p/\bar{\sigma})$ and $\rho = \tanh(\gamma\varepsilon_p/\bar{\sigma})$, with $\varepsilon_p = \varepsilon(t_1)$ being the fully integrated "area" of the pump function, and $\bar{\sigma} = \sqrt{|\sigma_r\sigma_s|}$. [For the opposite case $\sigma_r > 0$, $\sigma_s = -|\sigma_s|$, the imaginary $i$ is replaced by $-i$ in Eq.(25), but in this case $\hat{A}_s^{(+)}(t_1)$ and $\hat{A}_r^{(+)}(t_0)$ are identified as the inputs, rather than outputs.] After scaling the operators by their relative group slownesses by defining $\hat{B}_r^{(+)} = \sqrt{|\sigma_r|}\,\hat{A}_r^{(+)}$, $\hat{B}_s^{(+)} = \sqrt{|\sigma_s|}\,\hat{A}_s^{(+)}$, this operator transformation becomes

$$\begin{pmatrix} \hat{B}_r^{(+)}(t_0) \\ \hat{B}_s^{(+)}(t_1) \end{pmatrix} = \begin{pmatrix} \tau & i\rho \\ i\rho & \tau \end{pmatrix} \begin{pmatrix} \hat{B}_r^{(+)}(t_1) \\ \hat{B}_s^{(+)}(t_0) \end{pmatrix} \quad (26)$$

This is equivalent to a beam splitter transformation, with $\tau$ and $\rho$ playing the roles of transmission and reflection coefficients, respectively (with $\tau^2 + \rho^2 = 1$). This transformation is unitary, consistent with the discussions above for envelope reversal. The fact that a beam splitter transformation does not mix creation and annihilation operators shows that there is no added (spontaneous) noise, as there would be in amplification by phase conjugation or two-mode squeezing.

Accounting for free propagation from the input at $z = 0$ to the output at $z = L$ gives (using the diagram entry and exit times described above)

$$\hat{B}_r^{(+)}(L, -Mt - t_r) = \tau\hat{B}_r^{(+)}(0, -Mt) + i\rho\hat{B}_s^{(+)}(0, t)$$
$$\hat{B}_s^{(+)}(L, t + t_s) = i\rho\hat{B}_r^{(+)}(0, -Mt) + \tau\hat{B}_s^{(+)}(0, t) \quad (27)$$

where $t_r = |\sigma_r|L$, $t_s = |\sigma_s|L$. These can also be expressed as:

$$\hat{B}_r^{(+)}(L, t) = \tau\hat{B}_r^{(+)}(0, t + t_r) + i\rho\hat{B}_s^{(+)}(0, -t/M - t_s)$$
$$\hat{B}_s^{(+)}(L, t) = i\rho\hat{B}_r^{(+)}(0, -Mt + t_r) + \tau\hat{B}_s^{(+)}(0, t - t_s) \quad (28)$$

Note we used $t_r = Mt_s$. This shows that the transmitted components of the signal and idler are not time reversed, whereas the "reflected" (frequency converted) components, which are generated by the other mode, are time reversed.

A fruitful way to represent these input and output fields in terms of a complete set of TMs, $\{\varphi_j(t)\}$ is:

$$\hat{B}_s^{(+)}(0,t) = \sqrt{\sigma_s} \sum_j \hat{a}_j \varphi_j(t),$$

$$\hat{B}_r^{(+)}(0,t) = \sqrt{\sigma_r} \sum_j \hat{b}_j \varphi_j(-t/M) M^{-1/2}$$

$$\hat{B}_s^{(+)}(L,t) = \sqrt{\sigma_s} \sum_j \hat{c}_j \varphi_j(t-t_s),$$

$$\hat{B}_r^{(+)}(L,t) = \sqrt{\sigma_r} \sum_j \hat{d}_j \varphi_j(-t/M - t_r/M) M^{-1/2}$$

(29)

where $\hat{a}_j$ etc. are bosonic annihilation operators. Note the reversed time arguments in the representation of the $\hat{B}_r^{(+)}$ fields. Then Eq.(28) implies that the TM bosonic operators obey simple beam splitter relations for each pair of input modes:

$$\hat{d}_j = \tau \hat{b}_j + i\rho \hat{a}_j$$
$$\hat{c}_j = i\rho \hat{b}_j + \tau \hat{a}_j$$

(30)

Crucially, in the present case, the "beam splitter" coefficients $\{\tau, \rho\}$ are constant for all pairs of temporal modes, in contrast to the generic case where each TM pair experiences different coefficient values. [41]

To summarize the predictions of the analytical solution: If the pump function is short enough, an input pulse will be frequency converted and time reversed, and possibly stretched or compressed, with conversion amplitude (square-root of conversion efficiency) given by $\rho = \tanh(\gamma \varepsilon_p / \bar{\sigma})$. If $\rho = 1$, the entire complex pulse (including envelope and phase structure) of each input pulse is converted into a pulse at the other frequency, that is a swap operation $\omega_s \leftrightarrow \omega_r$ takes place, while inverting the direction of time within each.

The input-output relations derived above can be viewed as a "classical" mode transformation for temporal modes, analogous to the transformation effected by an optical beam splitter. If the initial state carries excitation in only one matched pair of input modes, $\varphi_j(t)$ for $s$ and $\varphi_j(-t/M)$ for $r$, then only one pair of output modes are excited, at least in the ideal limit considered in this paper. In this limit, therefore, we can use standard beam splitter theory [42] to transform the joint state of the input mode pair into the joint state of the output mode pair. Arbitrary states, including mixed states, can be treated this way.

For example, if the $s$ mode begins in a single-photon state and the $r$ mode in vacuum, i.e. $|1\rangle_s |0\rangle_r$, the output fields are in a superposition $\tau|1\rangle_s|0\rangle_r + i\rho|0\rangle_s|1\rangle_r$, where the $s$ and $r$ labels indicate central frequencies and matched-pair temporal mode shapes. Coherent-state inputs will transform as usual: $|\alpha\rangle_s |\beta\rangle_r \rightarrow |\tau\alpha + i\rho\beta\rangle_s |\tau\beta + i\rho\alpha\rangle_r$. An intriguing example is the interference of two single-photon states of different frequencies and temporal modes. Previous studies considered the generic case where photons of different color, but different, carefully matched TMs, experience a Hong-Ou-Mandel (HOM) second-order interference effect. [43] In the present case, we predict that if a "red" photon of an arbitrary TM and a "blue" photon in the corresponding time-reversed TM interact in a sum-frequency device having coefficients $\tau = \rho = 1/2^{1/2}$, both photons will emerge in the original TM or both will emerge in the time-reversed TM.

## 6. Numerical Simulations

How do the predictions made above bear up against "exact" numerical solutions of Eq.(22)? Figure 2 shows the outputs of four numerical simulations with the $A_r(0,t)$ pulse having zero amplitude at the input (along the top edge of each plot, where $z = 0$). For the numerical results, time is scaled by an arbitrary time $\tau_0$, and distance is scaled by a length $\tau_0 / \beta_p'$. All plots are given as positive amplitudes (absolute values) in the moving frame of the Gaussian pump, $A_p(z,t) = A_0 \exp(-t^2/P^2)$, which appears as a narrow spike at the center of Fig. 2(a) and is not shown in the other plots. The pump amplitude half-duration at $e^{-1}$ height is $P = 0.15$, and its peak amplitude is $A_0 = 1$, giving pulse area $\varepsilon_p = 0.266$. For Figs. 2 (a, b, d), the coupling coefficient has value $\gamma = 12$. For Figs. 2 (a, b, c), the group slownesses relative to the pump are $\{\sigma_s, \sigma_r\} = \{0.5, -0.5\}$. Figure 2(a) also shows a gaussian input pulse $A_s(0,t)$ (dark-shaded) with frequency $\omega_s$ propagating slower than the pump until entering the interaction window from the left at $z = 0$, then "reflecting" off the pump pulse to create the new pulse $A_r(z,t)$ (light-shaded) with frequency $\omega_r$ propagating faster than the pump (thus moving to the left). With the chosen parameter values, the frequency conversion is complete, consistent with the values of $\tau = sech(\gamma \varepsilon_p / \bar{\sigma})$, $\rho = \tanh(\gamma \varepsilon_p / \bar{\sigma})$, which evaluate to $\tau = 0.0034$, $\rho \approx 1.00$.

Figure 2(b) shows an asymmetric, nongaussian input pulse $A_s(0,t)$, which changes sign (phase) at the point where the amplitude touches zero. The frequency-converted envelope $A_r(z,t)$ is seen to be time reversed relative to the input pulse, as expected.

Figure 2(c) shows the same asymmetric input pulse $A_s(0,t)$, with the coupling coefficient reduced to a value $\gamma = 1.74$, chosen such that $\tau \approx \rho \approx \sqrt{1/2}$, so the expected frequency conversion efficiency is 0.5. Indeed, the converted and unconverted pulses are seen to have equal amplitudes at the output, as predicted.

Figure 2(d) shows the same asymmetric input pulse $A_s(0,t)$, with the coupling coefficient again increased to $\gamma = 12$, but now the group slownesses are chosen as $\{\sigma_s, \sigma_r\} = \{0.5, -0.25\}$, so the $A_r(z,t)$ pulse travels twice as fast (in the opposite direction in the pump frame) as the $A_s(z,t)$ pulse. The analytical theory above predicts a magnification factor of $M = |\sigma_r / \sigma_s| = 0.5$, and this is precisely what is seen in the figure: the converted pulse has half the duration as the input pulse.

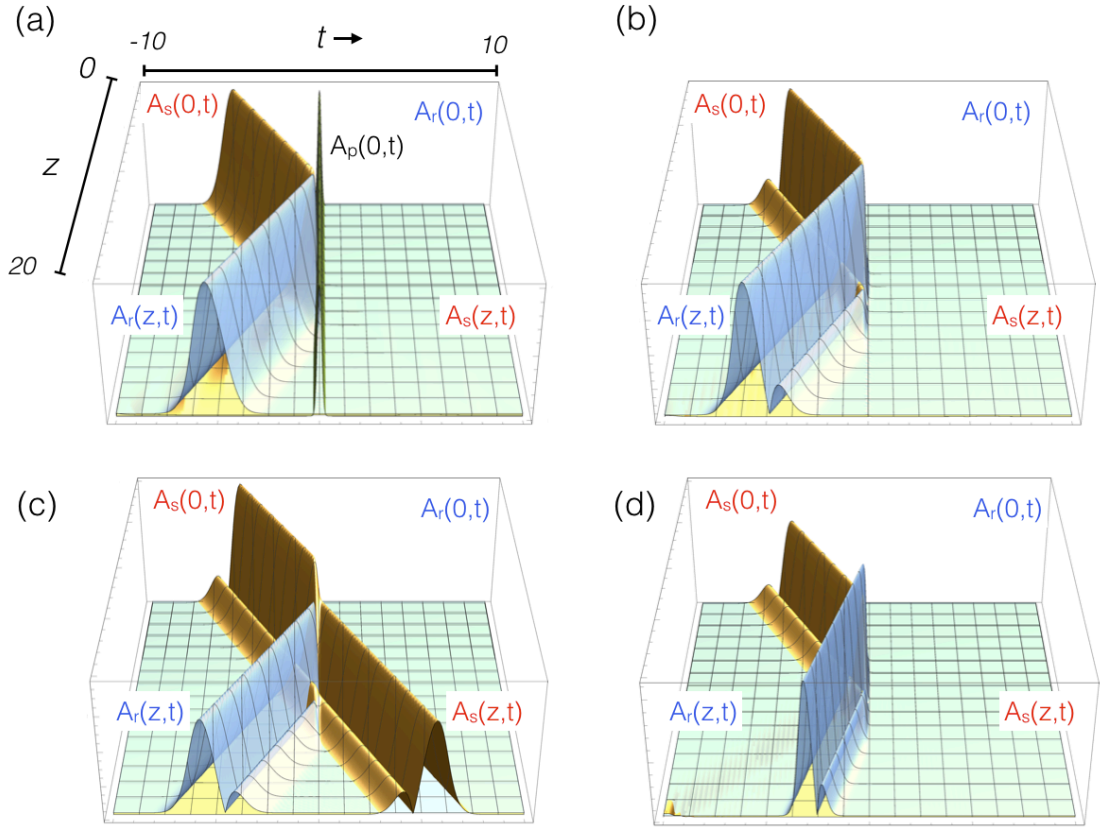

Figure 2 Amplitudes of pump, shown only in Fig. (a) at local time $t=0$, and two signal fields versus local time and distance traveled in the medium. (a) A Gaussian input is converted to a Gaussian output. (b) An asymmetric input is converted to a time-reversed replica. (c) With lower pump power the input is converted 50% to a time-reversed replica. (d) With unequal signal velocities, the converted replica is temporally compressed relative to the input. Throughout, the time span is 10 units, the length of the medium is 20 units, and the input location is taken to be at $z=0$.

The above examples considered input pulses with constant phase. The analytical theory predicts that for pulses with time-varying phases, the phase evolution will also be time reversed along with the (positive) envelope (although, again, the carrier wave is not phase conjugated). Consider an input pulse having the same envelope function as that in Fig. 2, but with a phase evolution arbitrarily chosen to have a quadratic time dependence, which corresponds to the linear spectral chirp that develops during propagation in a typical transparent medium such as an optical fiber:

$$\phi(t) = -0.15\, t^2 \tag{31}$$

Using the same parameters as in Fig. 2 yields the results in Fig. 3, where the envelopes and phases of the wave packets at the input and output of the medium are shown. If the output wave packet is flipped in time and delayed to match up with the input packet, both phase profile and envelope are found to match essentially perfectly (other than the numerical spikes in the output phase profile). This confirms the prediction of the analytical theory.

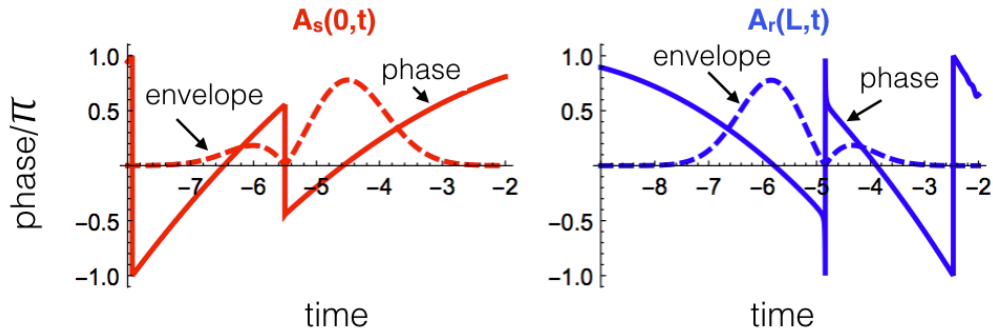

Figure 3 Envelopes and phases of input "red" signal field and output "blue" field versus time. Both envelope and phase are seen to be time reversed. The phase values are normalized by a factor $\pi$. (The unit of the envelopes is arbitrary.)

Our final example is an input envelope in the form of an exponentially decaying envelope, modeled as a smoothed version of $\exp(-t)\Theta(t)$, where $\Theta(t)$ is the Heaviside theta (step) function. Figure 4 shows this input envelope as $A_s(0,t)$ and the frequency-converted output envelope $A_r(z,t)$. The input pulse has been time shifted to view it in the same plot. Time reversal is seen clearly.

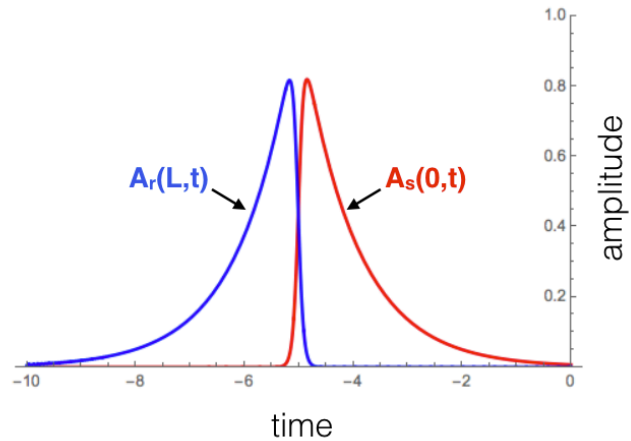

Figure 4 Envelopes of input "red" signal field and output "blue" field versus time. The input is shown delayed, and the output envelope is seen to be time reversed.

**7. Discussion**

Nonlinear wave mixing using very short pump pulses in a nonlinear-optical medium having suitable dispersion and group-velocity relationships can implement "blind" temporal envelope reversal of weak optical wave packets. The process can enable efficient transfer of information between quantum memories, which is optimized by "blind" envelope reversal of optical pulses with unknown temporal shape and unknown quantum state. Both shape and state must, of course, be within the operating domain (frequency band, temporal duration, and number of photons) of the device. In Appendix A we verified it is possible to have the needed relation of group slownesses $\beta_s' > \beta_p' > \beta_r'$ for $\omega_s < \omega_p < \omega_r$ using common wavelengths and a realistic second-order or third-order nonlinear optical medium. That is, we consider both sum-frequency generation or four-wave Bragg scattering, both of which are in-principle noiseless.

The predictions of the analytical solutions are found to hold in the numerical examples shown: If the pump function is much shorter than typical evolution times in the signal, and the group slowness conditions are met, then an arbitrary and unknown signal pulse will be time reversed (including envelope and time-varying phase structure), while being frequency converted, and possibly stretched or compressed. The conversion efficiency of the process is given by $\rho^2 = \tanh^2\left(\gamma \varepsilon_p / \sqrt{|\sigma_r \sigma_s|}\right)$, where $\gamma$ is the nonlinear coupling, $\varepsilon_p$ is the integrated pump amplitude, and $\sigma_j$ are the relative group slowness of the signal channels *s* and *r*. The combination $\gamma \varepsilon_p / \sqrt{|\sigma_r \sigma_s|}$ is a measure of the strength of interaction, which when large drives the conversion efficiency to unity. The overall phase is not conjugated, consistent with the process being unitary and noiseless in principle.

In practice, limitations will arise for signals having too-sharp features, necessitating extremely short pump pulses, eventually running into higher-order dispersion, which our model analysis omits. In applications, each specific realistic system should be carefully modeled to ensure the time reversal operation is faithful.

Our envelope-reversal result extends the possibilities for carrying out coherent quantum operations in a TM basis of single-photon states. For example, a "blind" temporal parity sorter would sort wave packets according to their TM parity under envelope reversal, that is, $\varphi_j(z,-t) = \pm\varphi_j(z,t)$, without knowing the wave packet. Such a TM parity sorter is an extension of transverse-spatial parity sorters, which have applications in quantum information science. [44,45,[46]] Details of a TM parity sorter using a two-stage envelope-reversal scheme are given in Appendix C.

Frequency conversion can also be accomplished using four-wave mixing in a third-order nonlinear optical medium. [25,47] It is easy to see that this mechanism could, in principle, also implement envelope reversal, if the two pump pulses are coincident and sufficiently short in duration, and if a medium is found that supports the necessary group-velocity relationships.

**Acknowledgements**

We thank Brian J Smith for helpful discussions. MR and DR were supported in part by NSF Grants QIS PHY-1521466 and AMO PHY‑1406354.

**APPENDIX A**

Here we verify it is possible to have group slownesses ordered as $\beta_s' > \beta_p' > \beta_r'$ (or the reverse order) in realistic second-order or third-order nonlinear optical medium. The material dispersion profiles of bulk second-order and third-order nonlinear optical media tend to have a monotonic relation between group-velocity and wavelength when the wavelengths are far away from zero-dispersion points or absorption peaks. This guarantees the required group-velocity relation for time-reversal, as long as the pump wavelength is in between the signal and idler wavelengths. The disadvantage of bulk media is that the wavelengths would either have to be far apart, or at least one of them would have to be close to an absorption peak, in order to maximize the relative difference of the group-velocities with respect to that of the pump pulse. Failing this, the media would have to be made substantially long in order to time-reverse temporally lengthy input pulses.

For example in Figs. 5 and 6, we consider bulk, periodically-poled lithium niobate as the medium. We consider wavelengths $\lambda_s = 1500\ nm$ and $\lambda_p$ in the range 700-900 nm, which puts $\lambda_r$ in the range 477-562 nm. In this range, while maintaining quasi-phase matching by suitable spatial poling periods (in the range 5-9 $\mu m$), the temporal demagnification is found to be, using realistic dispersion data, in the range 2 to 3. The maximum signal-input envelope width that would be time-reversed ranges from 22 ps to 15 ps in such waveguides of length 2.5 cm, with the limit growing linearly with length. Achieving a broader range of magnifications requires different media or wavelength choices.

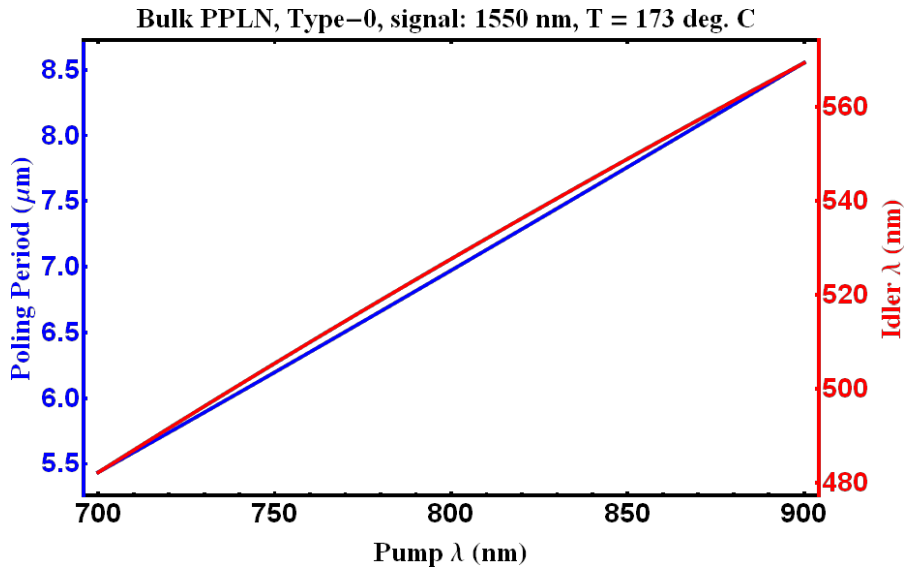

Figure 5 The poling period necessary for Type-0 phase matching using various pump wavelengths for sum-frequency generation of a signal at 1550 nm in bulk, periodically-poled lithium niobate (PPLN). Also shown are the corresponding idler wavelengths.

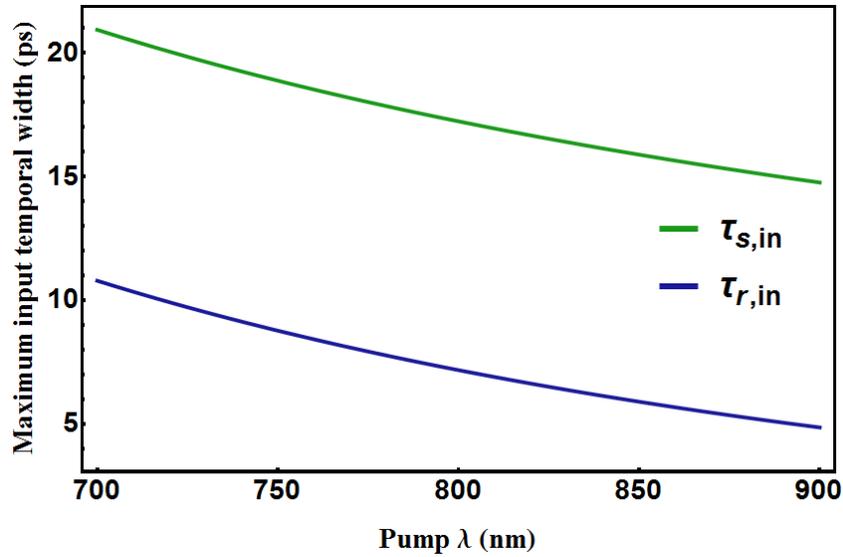

Figure 6 The maximum time durations of pulses in the signal (s) and idler (r) bands that would be time-reversed by a short pump pulse of various central wavelengths in a 25 mm long PPLN crystal that is poled for SFG of a 1550 nm signal pulse (see Fig. 5).

A means of overcoming this limitation is to use novel techniques like photonic crystals to introduce waveguide dispersion. For example, we consider the photonic crystal fiber used for frequency conversion using four-wave-mixing Bragg scattering in [48], which has been engineered to have zero-group-velocity-dispersion points at two wavelengths, as shown in Fig. 7. Such waveguides do not have monotonic relations between group-velocity and wavelength.

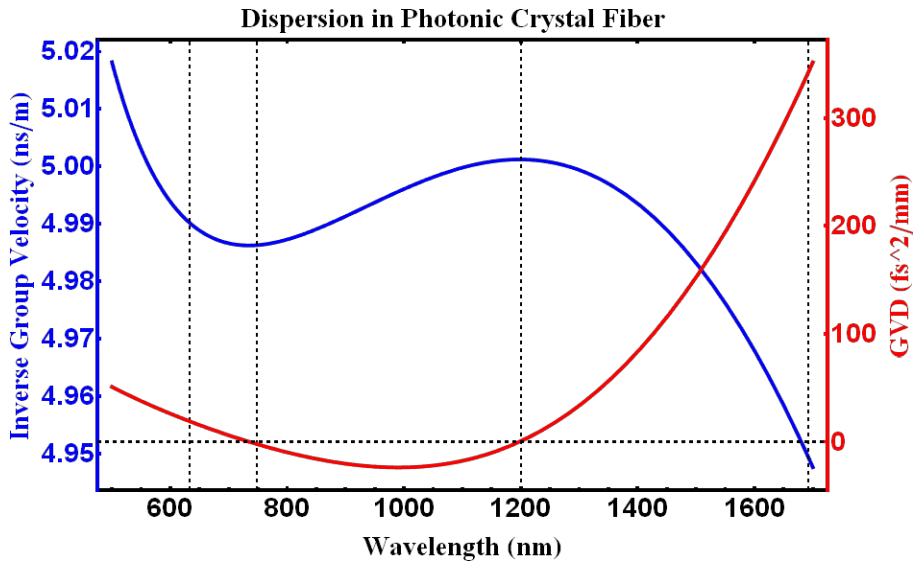

Figure 7 Inverse-group velocity and group-velocity dispersion (GVD) plots for the photonic crystal fiber used for frequency conversion via four-wave mixing in [48]. Of note are the two zero-dispersion points near 740 nm and 1200 nm.

The particular fiber from Fig. 7 is phase-matched for frequency conversion through Bragg-scattering [48] between the 1200 nm band and the 1692.4 nm band when pumped with strong, coherent lasers at 632.8 nm and 747.5 nm. Thus, when pumped asymmetrically [49] with continuous wave (or temporally long) pulses of helium-neon laser light at 632.8 nm and ~10 ps pulses of, say, Ti:sapphire laser at 747.5 nm, a fiber of length 100 m can time reverse a 3.7 ns long pulse centered at 1692.4 nm to create a 1.5 ns pulse in the 1200 nm band. These temporal durations are more conducive to coupling into known quantum optical memory systems than are the ps-duration pulses considered earlier. Careful dispersion engineering can be employed to modify the signal and idler wavelengths into more suitable bands such as telecom (1550 nm).

**APPENDIX B**

Here we verify the action of the monochromatic annihilation operator on a coherent state of a temporal mode, relevant to Eq. (19), and verify the form of the output state given in Eq. (19). Given a coherent state $|\alpha\rangle_1$ of a TM, defined by its creation operator:

$$\hat{a}_1^\dagger = \frac{1}{2\pi} \int_\Omega d\omega \, \tilde{\phi}_1(\omega) \hat{a}^\dagger(\omega) \;, \tag{32}$$

such that

$$\hat{a}_1 |\alpha\rangle_1 = \alpha |\alpha\rangle_1 \;. \tag{33}$$

A complete orthonormal set $\{\phi_j(\omega)\}$ of TMs, including $\tilde{\phi}_1(\omega)$, can be constructed, satisfying

$$\frac{1}{2\pi} \int_\Omega d\omega' \, \tilde{\phi}_j^*(\omega') \tilde{\phi}_j(\omega') = \delta(\omega - \omega') \;. \tag{34}$$

Then

$$\hat{a}_j^\dagger = \frac{1}{2\pi} \int_\Omega d\omega \, \tilde{\phi}_j(\omega) \hat{a}^\dagger(\omega) \;. \tag{35}$$

The monochromatic annihilation operator can be expanded as

$$\hat{a}(\omega) = \sum_j \tilde{\phi}_j(\omega) \hat{a}_j \;. \tag{36}$$

Then

$$\hat{a}(\omega)|\alpha\rangle_1 = \sum_j \tilde{\phi}_j(\omega) \hat{a}_j |\alpha\rangle_1$$
$$= \alpha \tilde{\phi}_1(\omega) |\alpha\rangle_1 \tag{37}$$

We use this result, with the standard commutator, and inserting a normalization coefficient $N$, to derive $\hat{U}^{(1)} |\Psi\rangle_{in} = \varepsilon |\Psi\rangle^{(1)}$, where $\varepsilon = -iC\alpha/N$ and

$$\left|\Psi\right\rangle^{(1)} = N\int_{0_s}\frac{d\nu}{2\pi}\tilde{\phi}_p(\omega_p+[m-1]\nu)\tilde{\varphi}(\nu)\hat{a}_r^\dagger(\omega_r+m\nu)\left|vac\right\rangle_{s,r}\otimes\left|\alpha\right\rangle_p \tag{38}$$

Replacing $\nu \to -\nu$ gives Eq. (19). The normalization coefficient is given by

$$N^{-2} = \int_{0_s}\frac{d\nu}{2\pi}\left|\tilde{\phi}_p(\omega_p+[m-1]\nu)\right|^2\left|\tilde{\varphi}(\nu)\right|^2 \tag{39}$$

**APPENDIX C**

A TM parity sorter can be constructed in a two-stage implementation, where the output from a first envelope-reversal stage, operating at 50% conversion efficiency, is sent into a second stage (having the pulses delayed to set up a repeat collision). This will function much like a Mach-Zehnder interferometer. In fact, such a scheme has been demonstrated for continuous-wave signals [50], and considered for pulsed signals [27]. In the present case, it serves to interfere a pulse with its time-reversed image. Although the two pulses are at different carrier frequencies, the frequency conversion process causes them to interfere, as in two-color photon interference [43], leading to the final output from the second stage.

Assume the phase shift between all fields between stages is zero. Then, if the input pulse is symmetric (even parity) in time, the output of this two-stage setup will be all blue. And if the input pulse is antisymmetric (odd parity) in time, the output of this two-stage setup will be all red. Finally, if the input pulse is neither symmetric nor antisymmetric, the pulse will be decomposed into its orthogonal parity components.